\documentclass[twocolumn,showpacs,amsmath,aps,pra,amssymb,nofootinbib,superscriptaddress]{revtex4-1} 
\usepackage{amsmath,amssymb,bm}
\usepackage{graphicx}
\usepackage{epstopdf}
\usepackage{latexsym}
\usepackage[normalem]{ulem} 
\usepackage[usenames,dvipsnames]{color}
\usepackage{hyperref}
\begin{document}
\newcommand{\Com}[1]{{\color{red}{#1}\normalcolor}} 

\newcommand{\ns}{\mathcal{N}_{\mathrm{s}}}
\newcommand{\sinc}{\mathrm{sinc}}
\newcommand{\nb}{\mathcal{N}_{\mathrm{b}}}
\title{ Dynamics of an itinerant spin-3  atomic dipolar gas in an optical lattice}

\author{Petra  Fersterer}
\affiliation{The Dodd-Walls Centre for Photonic and Quantum Technologies, New Zealand}
\affiliation{Department of Physics, University of Otago, Dunedin 9016, New Zealand}

\author{Arghavan Safavi-Naini}
 \affiliation{JILA, NIST, and Department of Physics, University of Colorado, 440 UCB, Boulder, CO 80309, USA}
 \affiliation{Centre for Engineered Quantum Systems, School of Mathematics and Physics,
The University of Queensland, St Lucia, QLD 4072, Australia}
\author{Bihui Zhu}
 \affiliation{ITAMP, Harvard-Smithsonian Center for Astrophysics, Cambridge, MA 02138, USA}
   \affiliation{Department of Physics, Harvard University, Cambridge, MA 02138, USA}
    \author{Lucas Gabardos}
 \affiliation{Universit\'e Paris 13, Sorbonne Paris Cit\'e, Laboratoire de Physique des Lasers, F-93430, Villetaneuse, France}
   \affiliation{CNRS, UMR 7538, LPL, F-93430, Villetaneuse, France}
    \author{Steven Lepoutre}
 \affiliation{Universit\'e Paris 13, Sorbonne Paris Cit\'e, Laboratoire de Physique des Lasers, F-93430, Villetaneuse, France}
   \affiliation{CNRS, UMR 7538, LPL, F-93430, Villetaneuse, France}
   \author{L. Vernac}
 \affiliation{Universit\'e Paris 13, Sorbonne Paris Cit\'e, Laboratoire de Physique des Lasers, F-93430, Villetaneuse, France}
   \affiliation{CNRS, UMR 7538, LPL, F-93430, Villetaneuse, France}
      \author{ B. Laburthe-Tolra}
 \affiliation{Universit\'e Paris 13, Sorbonne Paris Cit\'e, Laboratoire de Physique des Lasers, F-93430, Villetaneuse, France}
   \affiliation{CNRS, UMR 7538, LPL, F-93430, Villetaneuse, France}
\author{P. Blair Blakie}
\affiliation{The Dodd-Walls Centre for Photonic and Quantum Technologies, New Zealand}
\affiliation{Department of Physics, University of Otago, Dunedin 9016, New Zealand}

 \author{Ana Maria Rey}
 \affiliation{JILA, NIST, and Department of Physics, University of Colorado, 440 UCB, Boulder, CO 80309, USA}
 
  \affiliation{Center for Theory of Quantum Matter, University of Colorado, Boulder, CO 80309, USA}

\begin{abstract}
    Arrays of ultra-cold dipolar gases loaded in optical lattices are emerging as   powerful  quantum simulators of the many-body physics associated with the rich interplay between long-range dipolar interactions, contact interactions, motion, and quantum  statistics.  In this work we report on our  investigation of the  quantum many-body dynamics of  a large ensemble  of bosonic  magnetic chromium atoms with spin  $S=3$  in a three-dimensional lattice as a function of lattice depth.    Using extensive theory and experimental comparisons we study the dynamics of the  population  of the different Zeeman levels and the total  magnetization of the gas  across the superfluid to the  Mott insulator transition. We  are able to identify two  distinct regimes: At low lattice depths, where atoms are in the  superfluid regime, we observe that  the spin dynamics  is strongly determined   by the competition  between particle motion,  onsite interactions and external magnetic  field gradients. Contact spin dependent interactions  help to stabilize the collective spin length, which sets  the total magnetization of the gas.  On the contrary, at high lattice depths, transport is largely frozen out. In this regime, while  the spin populations  are  mainly  driven by long range dipolar interactions, magnetic field gradients also play a major role in the total spin demagnetization. We find that dynamics at low lattice depth is qualitatively reproduced by mean-field calculations based on the Gutzwiller ansatz; on the contrary, only a beyond mean-field theory can account for the dynamics at large lattice depths. While the cross-over between these two regimes does not correspond to sharp features in the observed dynamical evolution of the spin components, our simulations indicate that it would be better revealed by measurements of the collective spin length. 
\end{abstract}
\maketitle
\section{Introduction}
Ultra-cold gases   provide an excellent platform to study strongly correlated out-of-equilibrium quantum matter. So far, a broad range of atomic, molecular, and optical systems \cite{Gross2017,Bloch2008,Bohn2017} including  trapped-ions \cite{Blatt2012,Garttner2017,Bohnet2016, Neyenhuise2017} polar molecules \cite{Yan2013,Hazzard2014b}, Rydberg atoms \cite{labuhn2016,Zeiher2017,Bernien2017,Barredo2018,Guardado2018}, magnetic atoms \cite{Lepoutre2018b,Baier2016,dePazMISF,depaz2013,depaz2016,lev2018,Bottche2019,Chomaz2018}, and  cavity QED arrays \cite{Norcia2018,Davis2019} have been used to realize quantum many-body systems with long-range interactions and to probe equilibrium properties and  out-of-equilibrium dynamics both in  pinned and itinerant  systems. 

 Magnetic atoms trapped in optical lattices  ~\cite{Lahaye2009} naturally form a quantum simulator for complex $S>1/2$ models due to the  exponential growth in Hilbert space \cite{Aharonov2007,Hallgren2013}. In these systems, the large number of spin degrees of freedom, as well as the ability to create lattices with itinerant particles, where both motional and interaction effects can not be neglected,  quickly limit  the  capability of current state-of-the-art numerical methods to tackle the complex quantum dynamics. The high level of control and tunability in these simulation platforms has already resulted in numerous pioneering experiments in three dimensional optical lattices. These include studies  of extended Bose-Hubbard models \cite{Baier2016}, and   spin lattice models  \cite{manfred2019} with   erbium (Er) atoms using  $S=6$ bosonic and $F=19/2$ fermionic isotopes respectively,  as well as the spin  dynamics of $S=3$  bosonic chromium ($^{52}$Cr) atoms  in both  the superfluid and the Mott insulator regimes \cite{depaz2013,depaz2016,dePazMISF,Lepoutre2018b}. 

In this work, we present experimental results together with an extensive numerical study of the spin dynamics  seen in  an array of bosonic $^{52}$Cr atoms in a 3D lattice. By tuning the lattice depth, we explore the itinerant regime, where spin dynamics and tunneling occur over similar timescales. This regime is expected to be the most relevant for quantum simulations, since the complexity is then such that exact calculations are intractable by classical computers. We focus on lattice depths that span the Mott-to-superfluid transition. The dynamics is initialized by  rapidly rotating  a fully polarized equilibrium state, initially pointing  along the magnetic field direction, to orient the atomic spins in a direction perpendicular to the magnetic field, corresponding to a coherent superposition of all the magnetic sublevels. The system is then left to freely evolve at a fixed lattice depth. A residual magnetic field gradient and both dipolar and contact interactions generate spin dynamics beyond  the expected  simple spin precession. Notably, the populations of the seven different Zeeman sublevels are observed  to evolve following the spin rotation in a non-trivial way. We also theoretically investigate the dynamics of the total  magnetization (i.e. collective spin length) of the gas which tends to decay as the system evolves. 

Through our theoretical analysis we find that the system dynamics falls into two general regimes of behavior. (i) At low lattice depths, where the system is in the superfluid regime, spin transport is important as the gradient field drives the magnetic sublevels to spatially separate. The spin dynamics happens in a way that  is strongly affected by the  onsite interactions. We develop a mean-field Gutzwiller model \cite{Kimura2005a,Yamashita2007a} to study this regime. The Gutzwiller model is able to qualitatively  describe the experimentally observed population dynamics as well as  the inhibition  of the demagnetization process due to the spin dependent onsite interactions. This is similar to the protection against demagnetization and the persistence of ferromagnetic textures induced by the spin dependent contact interactions. These behavior was recently observed in a Bose Einstein condensate in a dipole trap \cite{Lepoutre2018a}.(ii) In deeper lattices transport is inhibited as the system  enters the Mott-Insulator regime. Here the dynamics of spin populations are driven primarily by long range dipolar interactions. We use  a generalized discrete Truncated Wigner Approximation (GDTWA) \cite{Lepoutre2018b,Schachenmayer2015a,Schachenmayer2015b,Polkovnikov} to describe the dynamics in this frozen atom regime. Similar observations were reported before at deep lattices \cite{Lepoutre2018b}, however what is surprisingly observed in this study is that the spin population dynamics remains almost independent of the lattice depth  as the system approaches the Mott insulating regime and that we are able  to reproduce well the observed  dynamics using the GDTWA method over a broad range of  moderate depth lattices. 

The Gutzwiller method predicts a strong reduction of spin dynamics as the lattice is raised and crosses the transition to the Mott-insulating state. However, this is at odds  with the experimental data, where a pronounced signature of the underlying quantum critical point in the spin population dynamics  is absent. We attribute this to the quantum fluctuations which are not taken into consideration in the Gutzwiller treatment. Instead of the abrupt change obtained within the Gutzwiller approach, what is observed is a gradual change between the two previously studied regimes: a classical ferrofluid without lattices \cite{PhysRevLett.121.013201}, and a correlated spin model at large lattice depths \cite{Lepoutre2018b}.  We note that a clearer signature of the transition appears in the theoretically calculated spin length, which we expect could be measured in future experiments.  


The paper is structured as follows. We discuss the experimental system and introduce the corresponding generalized Bose-Hubbard model that describes the dynamics. 
We then introduce the Gutzwiller and the GDTWA methods and present the results of a second order perturbative treatment of the spin dynamics.  Next, we study the dynamics starting from a simplified two-site model, which can be simulated exactly. We use the two-site model to identify key processes and to also benchmark  the validity of the Gutzwiller approximation and the perturbative treatments.  We then  use the different  numerical approaches to model the experimentally observed population  dynamics at various lattice depths. We also study the dynamics of the collective spin length (i.e. the dynamical evolution of the the total magnetization). We use both the populations and magnetization dynamics to look for signatures of the superfluid to Mott insulator transition and to understand the underlying physics in the two regimes.
\begin{figure*}
	\includegraphics[width=0.8\textwidth]{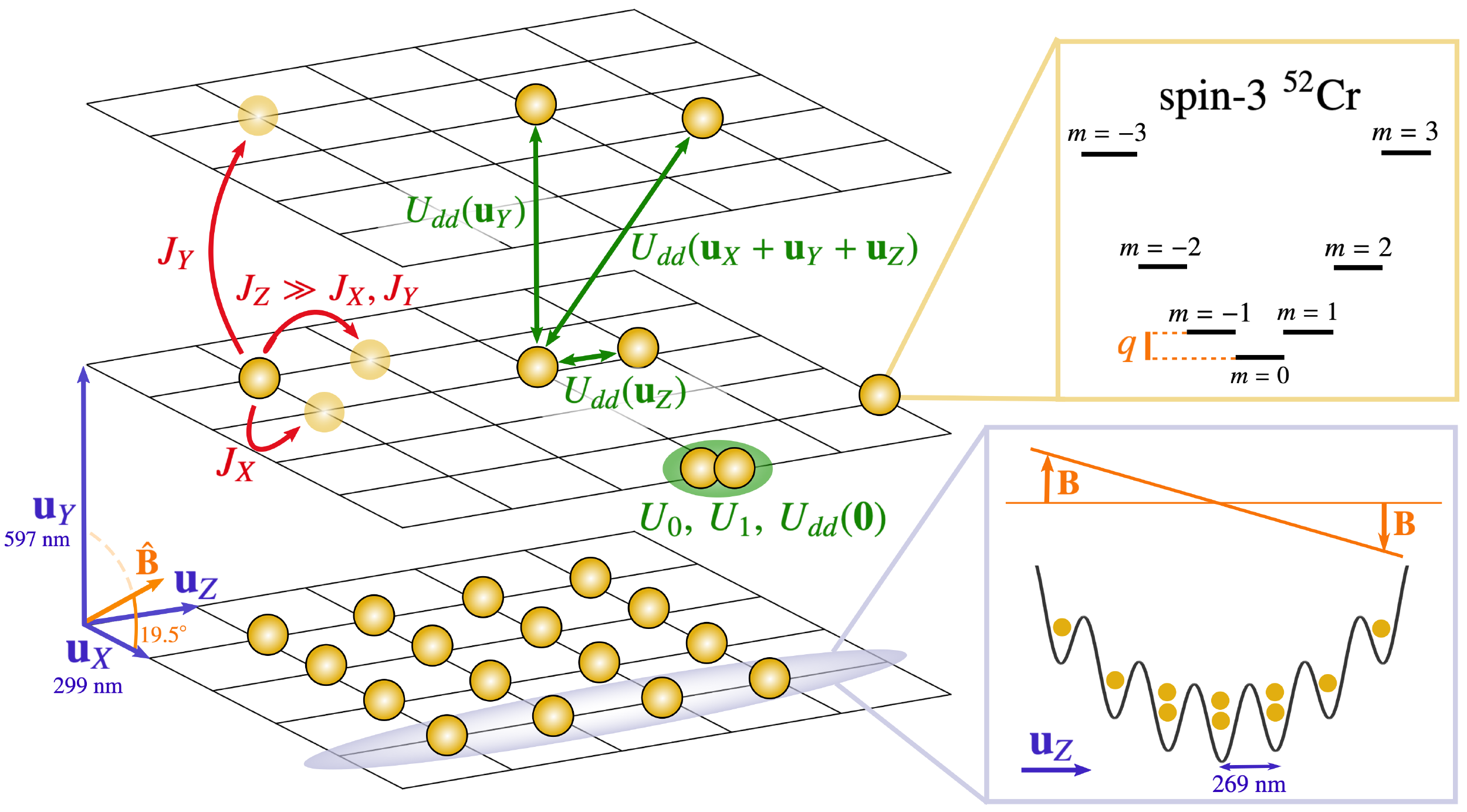}
	
	\caption{Schematic of system diagram indicating the 3D  geometry of the optical lattice, the interactions and tunnelling processes. The magnetic field (with direction $\mathbf{\hat{B}}$ in the $\mathbf{u}_X-\mathbf{u}_Y$ plane) has a gradient lying along the $\mathbf{u}_Z$ direction (lower inset).  Each  spin-3 chromium atom also has its sublevels shifted by a quadratic Zeeman term (upper inset).
	}
    \label{fig:exp_setup}
\end{figure*}

\section{System and Hamiltonian}

Chromium atoms are loaded into a three dimensional (3D) optical lattice as illustrated schematically in Fig.~\ref{fig:exp_setup}, and described by the lattice vectors $\{\mathbf{u}_X,\mathbf{u}_Y,\mathbf{u}_Z\}$. The lattice depth along each of the three lattice vectors is proportional to $V_0$.  The spatial extent of the system is determined by the harmonic confinement $\mathcal{V}_{\mathrm{tr}}=\frac{1}{2}m\sum_{\alpha=x,y,z}\omega_\alpha^2x_\alpha^2$, and the total number of atoms $N\approx 3\times10^4$. 
The spin degree of freedom is encoded in the Zeeman sublevels of the $S=3$ ground state of the ${}^{52}$ Cr atoms (see inset to Fig.~\ref{fig:exp_setup}). An external magnetic field in the direction $\hat{\mathbf{B}}$ sets the quantization axis, with the gradient in this field lying approximately along the $\mathbf{u}_Z$ direction (see inset to Fig.~\ref{fig:exp_setup}).

In the tight-binding regime Chromium atoms in the lowest band of the optical lattice are modeled as occupying Wannier state spatial orbitals $\mathrm{w}(\mathbf{r}-\mathbf{r}_i)$, centered on each lattice site $\mathbf{r}_i$. Here we consider the lattice site index $i=(i_X,i_Y,i_Z)$ a triad of integers, such that  $\mathbf{r}_i=\sum_{\alpha=X,Y,Z}i_\alpha\mathbf{u}_\alpha$.
The dynamics of our system is described by the generalized Bose-Hubbard model
\begin{align}
\label{eq:Ham}
\nonumber \hat{H}= & -\sum_{\alpha=X,Y,Z}J_{\alpha}\sum_{\langle i,j\rangle_{\alpha}}\sum_{m}\hat{a}_{m,i}^{\dagger}\hat{a}_{m,j} \\
\nonumber & +\sum_{i}\mathcal{V}_{\mathrm{tr}}(\mathbf{r}_i)\hat{N}_{i}\\
\nonumber & +q\sum_{j}\sum_{m}m^{2}\hat{N}_{mj} \\
\nonumber & -\gamma\sum_{i}b(\mathbf{r}_{i})\sum_{mn}S_{mn}^{z}\hat{a}_{m,i}^{\dagger}\hat{a}_{n,i}\\
\nonumber & +\frac{1}{2}\sum_{m,m',n,n'}\sum_{i}C^{mm'nn'}\hat{a}_{m,i}^{\dagger}\hat{a}_{m',i}^{\dagger}\hat{a}_{n',i}\hat{a}_{n,i} \\
 & +\frac{1}{2}\sum_{m,m',n,n'}\sum_{i,j} D_{ij}^{mm'nn'}\hat{a}_{m,i}^{\dagger}\hat{a}_{m',j}^{\dagger}\hat{a}_{n',j}\hat{a}_{n,i}
\end{align} 
where  the indices  $m, m', n, n'$ label the Zeeman sublevels of the atoms, and $\langle ..\rangle_\alpha$ is used to indicate nearest neighbours along direction $\mathbf{u}_\alpha$, with $J_\alpha$ being the tunneling amplitude in this direction. The operator $\hat a_{m,j}$  $\left(\hat a^\dagger_{m,j}\right)$ destroys (creates) a spin-3 bosonic particle in the $m$-Zeeman sublevel at site $j$,   $\hat{N}_{mj}=\hat{a}_{m,j}^{\dagger}\hat{a}_{m,j}$, and the total number of atoms on each site is given by $\hat{N}_{j}=\sum_{m}\hat{N}_{mj}$. 
The linear magnetic field gradient is described by $b(\mathbf{r})\approx b\,\mathbf{r}\cdot\mathbf{u}_Z$, with $S_{mn}^{x,y,z}$ being the $x,y,z$-components of the spin-3 matrices respectively and $\gamma=g\mu_{B}/\hbar$ where $g\simeq 2$ is the Land\'e-g factor and $\mu_{B}$
is the Bohr Magneton. 
The atoms also experience an effective quadratic Zeeman energy shift $q$ that arises from tensorial light shifts of the atomic levels. Both $q$ and $b$ vary with the lattice depth (e.g.~see Ref.\cite{Lepoutre2018b}).  

 
The last two terms in the Hamiltonian~\eqref{eq:Ham} describe the interactions. The contact interaction is of the form \cite{Kawaguchi2012a}
\begin{align}
C&^{mm'nn'}=  U_{0}\delta_{m,n}\delta_{m',n'}+U_{1}\sum_{\alpha}S_{mn}^{\alpha}S_{m'n'}^{\alpha},
 \label{Ham}
\end{align}
where $U_{n}=\tilde{c}_{n}\int d\mathbf{r}|\text{w}(\mathbf{r})|^{4}$. The contact interaction can also include higher order spin terms (for full form see Appendix \ref{app:parameters}), but for the states accessible during the dynamics in consideration  these have negligible effect compared to the contributions from the  $U_0$ and $U_1$ terms. Finally, the dipolar interactions are described by (see \cite{Kawaguchi2010a}) 
\begin{align}
\nonumber D_{ij}^{mm'nn'}&=\left(\frac{1}{2}S_{mn}^{x}S_{m'n'}^{x}+\frac{1}{2}S_{mn}^{y}S_{m'n'}^{y}-S_{mn}^{z}S_{m'n'}^{z}\right)\\
& \times U_{dd}(\mathbf{r}_{i}-\mathbf{r}_{j}),
\end{align}
where
\begin{align}
U_{dd}(\mathbf{R}) &  =\frac{4\pi}{3}c_{dd}\mathcal{F}^{-1}\left[\left(1-3\frac{(\mathbf{k}\cdot\hat{\mathbf{B}})^{2}}{k^{2}}\right)\mathcal{F}\left(|\text{w}(\mathbf{r})|^{2}\right)^{2}\right], \label{eq:Udd_def}
\end{align}
is the kernel of the time averaged dipole-dipole interaction between atoms in Wannier states separated by distance $\mathbf{R}$, with $c_{dd}=\mu_0(g\mu_B)^2/{4\pi}$.
Here we have used $\mathcal{F}$ and $\mathcal{F}^{-1}$ to denote the Fourier transform and its inverse, respectively. Because the dipole-dipole interactions decay rapidly with spatial separation, we restrict the summation to pairs of particles separated by up to one lattice site in each direction. This restriction reduces the computational difficulty, and we find that it does not affect the main results of our paper, as long as only a qualitative agreement is sought for. 

\begin{table}

\begin{tabular}{|c|c|c|c|}
\hline 
 & $V_{0}=3E_{r}$ & $V_{0}=9E_{r}$ & $V_{0}=15E_{r}$\tabularnewline
\hline 
\hline 
$U_{0}/h$ (Hz)  & $1250$ & $2860$ & $4190$\tabularnewline
\hline 
$\omega_{Z}/2\pi$  (Hz)& $279$ & $337$ & $387$\tabularnewline
\hline  
\hline 
$J_{X}/h$  (Hz)& $917$ & $106$ & $18.6$\tabularnewline
\hline 
$J_{Y}/h$  (Hz)& $11.8$ & $5.67\times10^{-2}$ & $1.11\times10^{-3}$\tabularnewline
\hline 
$J_{Z}/h$  (Hz)& $1380$ & $297$ & $82.3$\tabularnewline
\hline 
$U_{dd}(\mathbf{0})/h$  (Hz)& $-6.96$ & $-15.9$ & $-23.2$\tabularnewline
\hline 
$U_{dd}(\pm\mathbf{u}_{X})/h$  (Hz)& $2.86$ & $3.02$ & $3.02$\tabularnewline
\hline 
$U_{dd}(\pm\mathbf{u}_{Y})/h$  (Hz)& $-0.176$ & $-0.175$ & $-0.175$\tabularnewline
\hline 
$U_{dd}(\pm\mathbf{u}_{Z})/h$  (Hz)& $-2.52$ & $-2.60$ & $-2.62$\tabularnewline
\hline 
\end{tabular}\caption{Parameters, for different  lattice depths $V_0$ measured in units of the recoil energy $E_r$ for  $\lambda=532\,$nm.  $\omega_z$ is the estimated trapping frequency in the direction of the magnetic field gradient.  
 After the loading, sites are populated at the most with three atoms, which is the upper cutoff of our simulation. Note $U_1=7.40\times 10^{-2}U_0$, $U_2=0.795U_0$ and $U_3=-4.71\times 10^{-3}U_0$. The gradient, which does not vary with lattice depth, is given by $\gamma b(\mathbf{r}_i)=29\times 10^6 \text{ Hz m}^{-1} \times \mathbf{r}_i\cdot\hat{\mathbf{u}}_Z$ where the lattice spacing in the $\hat{\mathbf{u}}_Z$ direction is 269 nm. The values of the quadratic Zeeman shifts used in the simualtions are given in Fig.~\ref{fig:experiment_comp}(o)} \label{Experimental_Params}
\label{tab:params}
\end{table}

To understand the system dynamics it is important to quantify the many microscopic parameters of the system which appear in the Hamiltonian. We summarize these values for three cases of the lattice depth $V_0$ in Table~\ref{tab:params}. We note that the 3D lattice considered here is non-separable and we approximate the Wannier functions of the ground-band as Gaussians in order to obtain estimates of the interaction terms $\{U_0,U_{dd}(\mathbf{R})\}$.  The initial distribution of the atoms in the lattice depends on lattice depth because it varies the strength of the onsite interactions ($U_n$) but also because the focused lasers used to make the lattice contribute to the harmonic confinement which increases with lattice depth.  

In our experiments an initial state is produced by loading a Bose-Einstein condensate of Cr atoms spin polarized in the $m=-3$ sublevel into the optical lattice at the desired depth $V_0$. We assume that the atoms are initially in equilibrium. For our parameters the superfluid to Mott insulator transition is predicted to occur according to the Gutzwiller model at approximately $V_c\approx 8E_r$, where $E_r$ is the recoil energy for $\lambda=532\,$nm. A fast  $\pi/2$ microwave pulse ($\sim 5 \mu$s long) is then applied to rotate the atomic spins to be along the spin-$x$ axis, and thus in a superposition of all $7$ Zeeman sublevels. The corresponding  populations  $N_m(t)$ are then measured as a function of hold time after the pulse, using Stern-Gerlach imaging. The observed evolution in these populations (see data on Fig.~\ref{fig:experiment_comp}) motivates the theoretical analysis that we develop in the following section.



\section{Theoretical methods}
\subsection{Gutzwiller method}\label{SecGutz}

 The Gutzwiller method is a meanfield technique suited to describing bosons in an optical lattice. This approach has been applied to spin-1 bosons (e.g.~see  \cite{Kimura2005a,Yamashita2007a}), but as far as we are aware our work is the first to consider the extension of this method to describe spin-3 bosons with dipole-dipole interactions. The Gutzwiller method treats onsite terms exactly and inter-site couplings (due to tunneling and interactions) at the meanfield level. 
 The system state in this method is written as a product of states at each lattice site, i.e.~
\begin{align}
\vert \psi(t)\rangle_{G}&=\prod_{i}\vert\psi(t)\rangle_{i},\\
\vert\psi(t)\rangle_{i}&=\sum_{\mathbf{N}}f_{\mathbf{N}}^{i}(t)|\mathbf{N}\rangle_{i},\label{psiidef}
\end{align}
where $\vert \mathbf N \rangle_i =\vert N_{3},N_{2},N_{1},N_{0},N_{-1},N_{-2},N_{-3} \rangle_i$ is the spin-3 Fock state basis at site $i$ and $f_{\mathbf{N}}^{i}$ are the respective amplitudes of the onsite expansion. It can then be shown that the equation of motion for the state at site $i$ is given by
\begin{align}
i\hbar\frac{d}{dt}|\psi(t)\rangle_{i}=\hat{H}_{G}^{i}|\psi(t)\rangle_{i},\label{motion}
\end{align}
where $H_{G}^{i}$ is the Hamiltonian at lattice site $i$. This contains the exact onsite Hamiltonian $\hat{H}_{i}$ with all the onsite terms of Eq.(\ref{eq:Ham}), and the mean field contribution from nearest neighbours, giving  \cite{Jreissaty:2011id} 
\begin{align}
	\hat{H}_{G}^{i} &=	\hat{H}_{i}\label{HG}\\ 
	\nonumber &-\sum_{\alpha}J_{\alpha}\sum_{j^{(i)}_{\alpha}}\sum_{m}\left(\langle\hat{a}_{m,j}\rangle\left[\hat{a}_{m,i}^{\dagger}-\frac{1}{2}\langle\hat{a}_{m,i}^{\dagger}\rangle\right]+h.c.\right)\\
	\nonumber	&+\sum_{m,m',n,n'}\sum_{j\neq i}D_{ij}^{mm'nn'}\langle\hat{a}_{m',j}^{\dagger}\hat{a}_{n',j}\rangle\\
	\nonumber&\qquad\qquad\qquad\qquad\quad\times\left(\hat{a}_{m,i}^{\dagger}\hat{a}_{n,i}-\frac{1}{2}\langle\hat{a}_{m,i}^{\dagger}\hat{a}_{n,i}\rangle\right),
\end{align}
where $j^{(i)}_{\alpha}$ are the nearest neighbours to site $i$ in the $\alpha$ direction, and we also choose to only include the sum over nearest neighbour $j$ in the DDI term. 

The Gutzwiller ansatz reduces the Hilbert space dimension significantly and couples sites only through meanfield terms. As such, if every site is identical, the 3D problem only requires the solution at one site, i.e.~ $f_{\mathbf{N}}^{i}(t)\to f_{\mathbf{N}}(t)$ for all  $i$. However, in our problem the magnetic field gradient and the harmonic confinement break the translational symmetry. Thus, the system can no longer be reduced to a single site within the Gutzwiller approximation. To produce a computationally tractable case we use a model with spatially varying coefficients only  along the $\mathbf{u}_Z$ direction and assume the other two  perpendicular  directions to be spatially invariant, i.e.~we use the ansatz  $ f_{\mathbf{N}}^{i_Z}(t)$ to  describe the dynamics which assumes the sites $i=(i_X,i_Y,i_Z)$ behave identically to  $(0,0,i_Z)$. This choice leaves us explicitly modelling an effective  one-dimensional problem along the direction of the gradient field, as schematically shown in the inset to Fig.~\ref{fig:exp_setup}, yet allows us to retain the 3D character of the long range DDIs and tunnelling.

Due to the spin-degrees of freedom the local Hilbert space grows rapidly with the number of atoms per site. We restrict  $f_{\mathbf{N}}^{i_Z}$ to $\{\mathbf{N}: \sum_mN_m\le 3\}$, i.e.~up to 3 atoms per site, which should be adequate as the experimental initial state typically has some occupation of doublons (i.e.~sites with 2 atoms), but the number of triply occupied sites is negligible both initially and during the dynamics. To match the experimental system we consider a system of 30 lattice $i_Z$-sites (see Appendix \ref{app:nummethods} for additional details). The initial state is computed by finding the Gutzwiller ground state for $N_{\rm tot}=18$ atoms (occupying the 1D line of sites along the gradient direction), all restricted to the $m=-3$ sublevel.
We note that this choice for $N_{\rm tot}$ does not reproduce the initial number of doublons observed experimentally. However, in the experiment the total number of atoms is observed to decrease rapidly with time as  $N(t)\approx N(0)\exp(-\gamma t)$, where $\gamma\approx 100 s^{-1}$ (see Appendix \ref{app:parameters}) due to dipolar relaxation. Our Gutzwiller  theoretical model does not include losses, but our choice of $N_{\rm tot}$ gives a doublon fraction closer to the average value observed experimentally over the $\lesssim10\,$ms time period. This accounts for the rapidly decaying population of doubly occupied sites and our effective trapping conditions(the choice of $N_{\rm{tot}}$ is discussed further in Sec.~\ref{sec:fulldynamics}). The initial state density distribution for various lattice depths is shown as the black line in Fig.~\ref{fig:experiment_comp}(p), with the result for $V_0=15\,E_r$ demonstrating a uniformly filled Mott insulator state.  
To initiate the dynamics we apply a $\pi/2$ spin rotation about the $y$ axis to this initial state.


\subsection{GDTWA}
\text

In the deep lattice regime we describe the system with a spin Hamiltonian of the form:
\begin{align}
\hat H&=\sum_i b({\bf r}_i)\hat S_i^z+q\sum_i  (\hat S_i^z)^2\nonumber\\
&+\frac{1}{2}\sum_{i,j\neq i}U_{dd}({\bf r}_i-{\bf r}_j)[\frac{1}{2}(\hat S_i^x\hat S_j^x+\hat S_i^y\hat S_j^y)-\hat S_i^z\hat S_j^z], \label{eq:twaH}
\end{align}
where the first two terms account for the linear and quadratic Zeeman fields and the last term describes the long-range dipolar interactions that couple Cr atoms  in different sites. A fraction of sites are doubly occupied in the experiment and we model each of these sites as a pseudo-atom of spin-6, {\em i.e.} the maximum total spin of two spin-3 Cr atoms, and their dynamics are still  governed by Eq.~(\ref{eq:twaH}), but with $\hat S^{x,y,z}$ replaced by operators corresponding to $S=6$. This treatment is motivated by the fact that the initial spin state for each doubly occupied site is in the symmetric $S=6$ manifold, and the dominant on-site contact interactions do not change this total spin~(see Eq.(\ref{Ham})). We have also checked numerically that this is a good approximation. We solve the spin dynamics under the Hamiltonian Eq.~(\ref{eq:twaH}) by applying the generalized discrete Truncated Wigner Approximation (GDTWA), a numerical approach first introduced in Ref.~\cite{Lepoutre2018b}. In the GDTWA approach, quantum dynamics is obtained by  proper sampling the initial quantum state in phase space and averaging over the ensuing trajectories \cite{Erbium2019}. The GDTWA method has been shown to be capable of capturing quantum correlations developed during spin dynamics~\cite{Lepoutre2018b,manfred2019}. With this method, we are able to perform calculations including the effect of quantum fluctuations in a relatively large system (  lattice size $13\times 6\times 13$ ) and check that a convergence in system size  is reached within experimental uncertainties. The population of a spin state $m$ is obtained via combining the contributions from both singly and doubly occupied sites
\begin{align}
N_m(t)&=\eta(t)\sum_{M=-6}^{6}\langle 3,m,3,M-m|6,M\rangle^2 N^{\rm S=6}_{M}(t)\nonumber\\
&+(1-\eta(t)) N^{\rm S=3}_m(t),
\end{align}
where $N_M^{S=3,6}(t)$ is the averaged population on a spin state $M$ calculated for the singly and doubly occupied sites, respectively. The term  $\langle 3,m,3,m'|6,M\rangle$ denotes the Clebsch-Gordan coefficients, and $\eta$ denotes the fraction of atoms in doubly occupied sites. Due to atom loss from doubly occupied sites, $\eta$ varies with time. The time dependence that we use for $\eta$ is  extracted from  experimental measurements  of the  total atom number $N(t)$ (see Appendix \ref{app:parameters}). 

\subsection{Perturbative treatment}

Due to the complexity of the Hamiltonian~\eqref{eq:Ham} many factors  play a role in the system evolution following the spin rotation.
 In order to have a better grasp of the contributions of the various terms in the Hamiltonian we first consider the dynamics of the system in the short-time limit using perturbation theory.

We consider the dynamics of an initial state with $N=2$ atoms occupying two neighbouring lattice sites, $i=1,2$, described by the initial state
\begin{align}
\nonumber \vert\psi(0)\rangle&=\sum_{m,n=-3}^{3}f_{m}f_{n} \hat a_{m, 1}^\dagger \hat a_{n, 2}^\dagger  \vert 0 \rangle_1 \vert 0 \rangle_2.
\end{align}
where $\vert 0 \rangle_i$ is the vacuum state in site $i$. The $f_{m}$ coefficients specify the different spin
amplitudes after the spin rotation, $f_{0}=-\frac{\sqrt{5}}{4}$ and  $f_{\pm1}=\frac{\sqrt{15}}{8}$ , $f_{\pm2}=-\sqrt{\frac{3}{32}}$, and $f_{\pm3}=\frac{1}{8}$. 
To second order in $t$ the evolution of the  population of the different Zeeman levels is given by:
\begin{align}
\nonumber N_{m}^{(2)}(t)=\langle &\hat{N}_{m}(0)\rangle\\
&-it \biggl\langle\left[\hat N_{m},\hat{H}_2\right]\biggr\rangle-\frac{t^{2}}{2}\biggl\langle\left[\left[\hat{ N}_{m},\hat{H}_2\right],\hat{H}_2\right]\biggr\rangle,
\end{align}
where $\hat{H}_2$ is the Hamiltonian (\ref{eq:Ham}) specialized to two sites separated by $\mathbf{u}_Z$.
 
 We find that the term proportional to $t$ vanishes, and to second-order in $t$, the spin  population dynamics for $m=0$ is given by:
\begin{align}  
N^{(2)}_{0} (t)&= \frac{ 5}{8}- \frac{45}{16}t^{2}U_{dd}(\mathbf{u}_Z)\left[q+\frac{3}{4}U_{dd}(\mathbf{u}_Z)\right]. \label{N0pert}
 \end{align} 
  The evolution of the  population of the other sublevels is of a similar form and  presented in Appendix \ref{app:pertexp}. We have also ignored in the perturbative treatment the dynamics of doubly occupied sites. As explained  in the Appendix \ref{app:timescale}, the perturbative treatment tends to break down very quickly when those terms are included and a better approximation is obtained when they are excluded. We also note that  up to this order, our perturbative result immediately generalizes to the many-body system by simply summing over all pairs of particles. 
 
\begin{figure*}
	\includegraphics[width=1\textwidth]{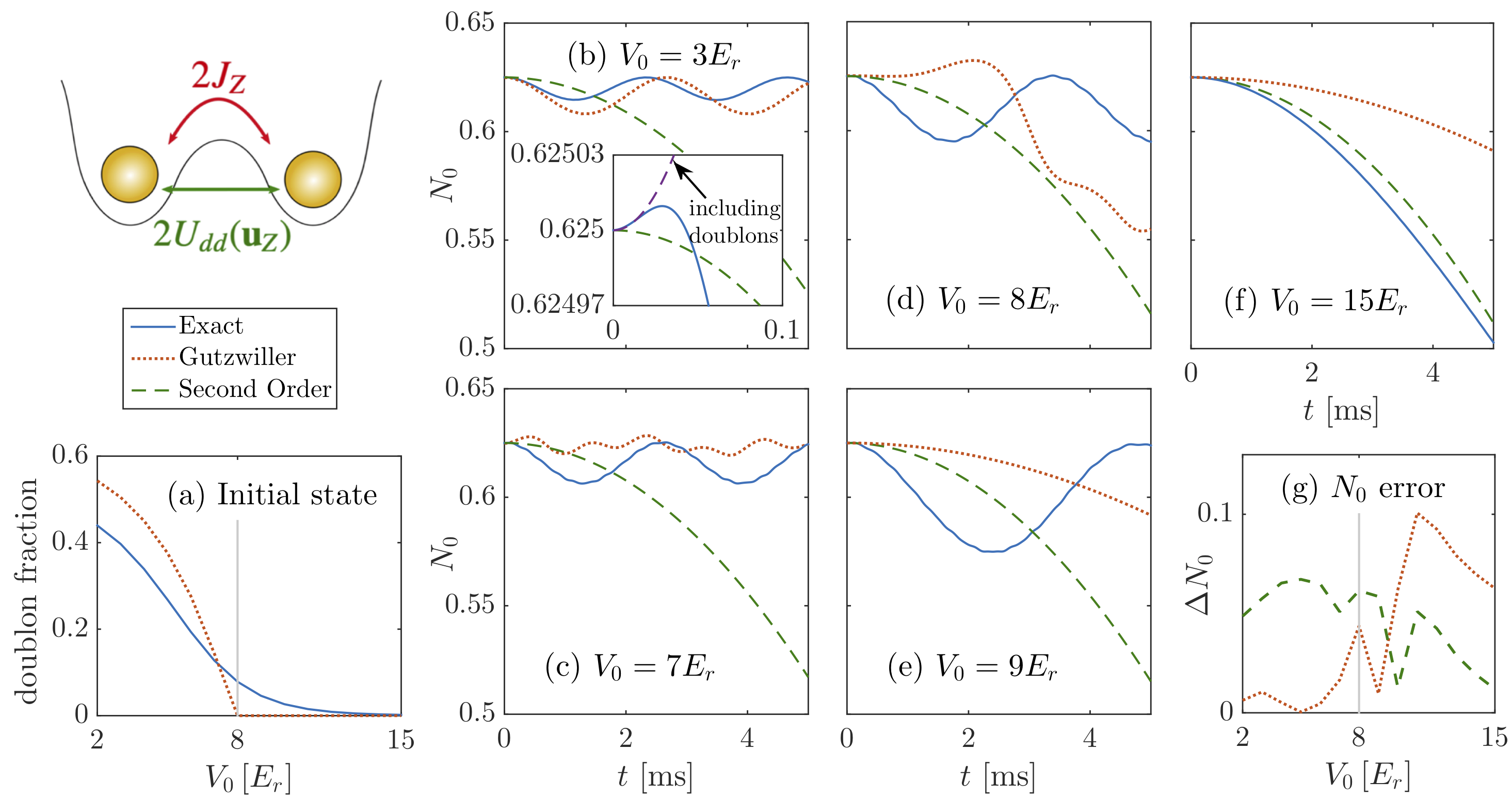}
	
	\caption{Comparison of predictions for $N_0$ population dynamics in the double-well system $\hat{H}_2$. (a) Doublon fraction of the two-particle initial state.
	(b)-(f) The population of the $m=0$ substate as a function of time for a double-well system allows us to compare the short-time perturbative expansions at second-order to the Gutzwiller dynamics, as well as the exact dynamics for five different lattice depths. In the deep lattice the second order formula is a good approximation for longer times.  The simulations parameters are given in Table \ref{Experimental_Params}, with the tunnelling and nearest neighbour DDIs adjusted  so the lattice depth at which the transition occurs is similar to the  3D system and  $q/h=2$ Hz. Inset in (b) shows the perturbative result (\ref{N0pert}) extended to include doublons, demonstrating that they give rise to rapid short time dynamics captured in the exact solution, but limiting the applicability of the perturbative result to short times. (g) Difference of Gutzwiller and second order populations compared to the exact population at 4 ms, as a function of lattice depth.}
    \label{fig:shorttime}
\end{figure*}

Equation (\ref{N0pert}) provides valuable insight into the short-time dynamics of the system. First, it shows that the dipole-dipole interactions between  neighboring sites and the quadratic Zeeman term  are  what drive the dynamics at short time for all lattice depths. Tunneling or linear magnetic field gradients,  when present,  only cause redistribution of the spin populations at  a higher order (See Appendix \ref{app:4thorder}).


\section{Analysis of Dynamics}

\subsection{Double-well Dynamics}

The computational complexity of simulating Hamiltonian~\eqref{eq:Ham} makes any exact treatment beyond a few lattice sites challenging. In order to benchmark our methods we compare the predictions of the Gutzwiller dynamics and the perturbative expressions obtained above to the exact dynamics of a double-well system with two atoms. In Fig.~\ref{fig:shorttime}(a) we plot the proportion of  doubly occupied sites (doublons) in the initial states. This shows that the Gutzwiller approximation predicts that the sites become single occupied for $V_0\gtrsim8 E_r$, indicating the existence of a  superfluid to Mott-insulator transition at $V_0\approx8\,E_r$. In contrast, the exact double-well result shows that the doublon fraction varies smoothly with lattice depth. Nevertheless, the range of lattice depths where doubly occupied sites are dominant is well captured by the Gutzwiller model. 
 The panels (b)-(f) compare the perturbative, exact and the Gutzwiller dynamics for the $m=0$ population.  
As a meanfield method the Gutzwiller approach is not expected to provide an accurate description of a two-particle system, however the comparison reveals that Gutzwiller results are qualitatively correct for low lattice depths $V_0\lesssim8 E_r$. In this regime the populations tend to oscillate. Furthermore, we note that we have not included the role of doubly occupied sites in the above expressions (see Appendix \ref{app:pertexp} for the full results). We find that by including the doublon sites the solution matches the initial curvature of the exact solution (see the inset of Fig.~\ref{fig:shorttime} (b)). However this leads to a divergence between the two approaches for longer times (see Appendix \ref{app:timescale}). The slower dynamics of the second order perturbation theory for singly occupied sites follow the exact solution for a longer time as shown throughout the panels. For deeper lattices $V_0>9 E_r$, where the system is in the insulating regime and doublons are negligible, the $N_0$ population tends to decay, and is well described by the perturbative result. In this regime on the contrary the Gutzwiller model  shows suppressed population dynamics. This reveals the failure of the   method  to  account for the necessary quantum fluctuations that mainly  drive the dynamics in the frozen atom  limit.

 To quantify the validity of the different methods, in panel (g) we plot the accumulated  error during the time evolution vs lattice depth. It explicitly shows that while the Gutzwiller gives a better description at low lattice depths, the perturbative formula  describes well the short time dynamics in the Mott regime. Note that the GDTWA results were not shown in this two site comparison as GDTWA is a method based on semi-classical trajectories, which is more valid with larger $N$. Therefore it does not work for the extreme case of only two atoms.

\subsection{Full system dynamics}\label{sec:fulldynamics}

\begin{figure*}
	\includegraphics[width=1\textwidth]{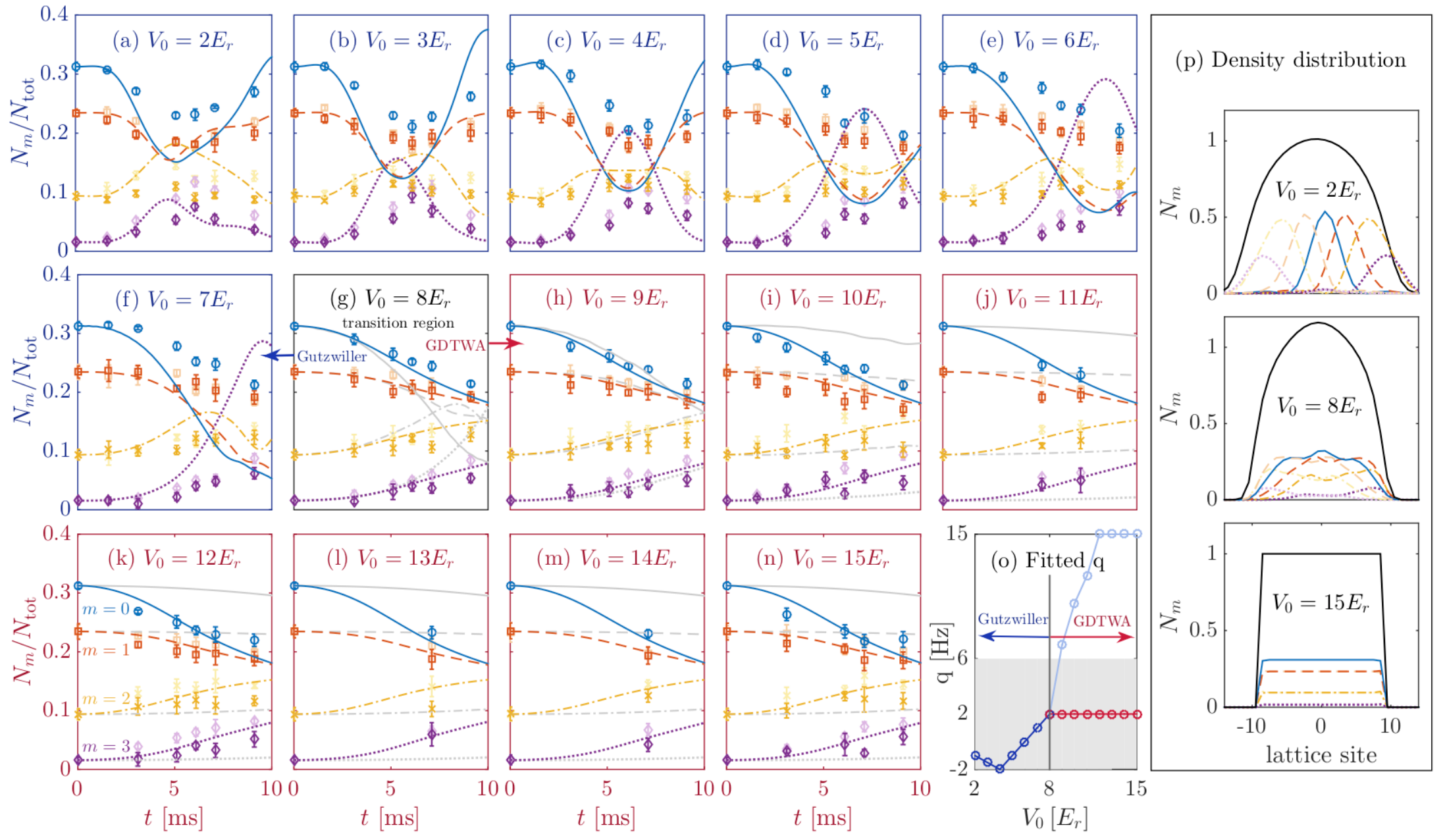}
	
	\caption{Comparison of simulation (lines) to experiment (markers). Blue (solid, circles) indicates $m=0$, red (dashed, square) indicates $m=1$, yellow (dot-dashed, cross) indicates $m=2$, and purple (dotted, diamond) indicates $m=3$. Pale markers indicate the negative $m$ experimental populations. For $V_0<V_c$ the Gutzwiller model is shown, while for $V_0\geq V_c$ the GDTWA model is shown, with the corresponding $q/h=2$ Hz Gutzwiller solution shown in grey.  Experimental error bars are from statistical standard errors and from 10$\%$ uncertainty in the estimated lattice depths. (o) The quadratic Zeeman field $q$ fitted at each lattice depth in the simulations based on the Gutzwiller ansatz, and $q/h=2$ Hz value used in the GDTWA. The light grey region indicates the range of acceptable $q$ values. The light blue circles correspond to optimal values for the Gutzwiller approximation which however are outside the acceptable range    (p) Distribution of whole cloud across the lattice at $t=0$ (black line) and distribution of individual Zeeman levels across the lattice at $t=5$ ms.} 
	\label{fig:experiment_comp}
\end{figure*}
\begin{figure*}
	\includegraphics[width=1\textwidth]{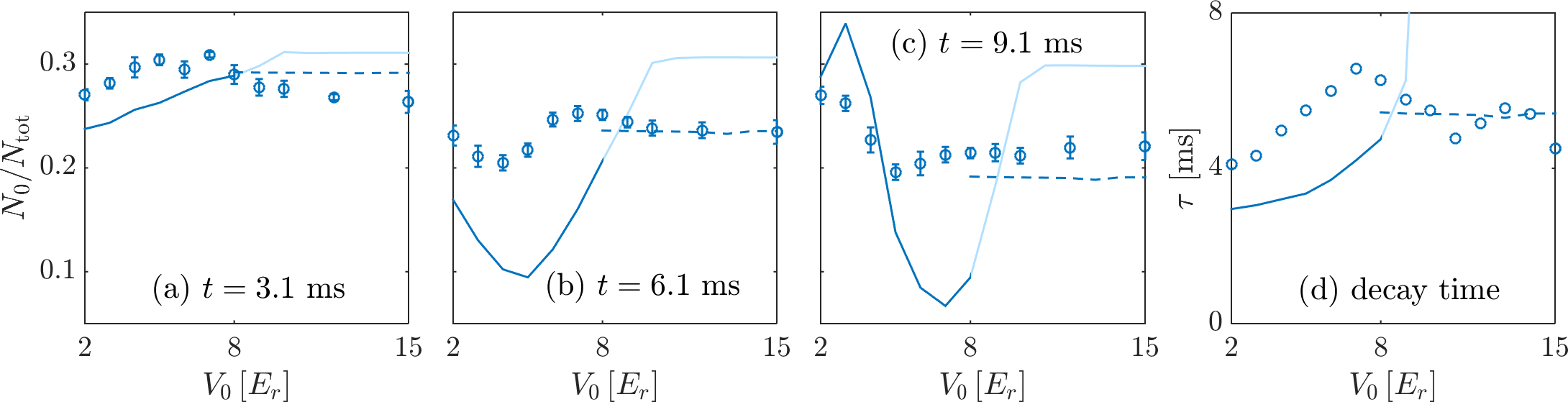}
	
	\caption{(a-c) Gutzwiller (solid line) and GDTWA (dashed line) $m=0$ populations compared to experimental results (symbols) as a function of lattice depth. (d) Time $\tau$ for $m=0$ population to decay to $N_0/N_{\rm{tot}}=0.25$.}.
    \label{fig:withV0}
\end{figure*}

 In this section we use the Gutzwiller ansatz and the GDTWA  to study the dynamics of larger systems in the   parameter regimes  used for the experimental measurements of the spin population dynamics. 

As  discussed above, in  the Gutzwiller model we  do not include the particle-loss (due to dipolar relaxation) dynamically. Instead, we account for it by sweeping over a range of particle numbers $N_{\rm tot}$ and find that the experimental dynamics are best captured by $N_{\rm tot}=18$.  
We note that the Gutzwiller treatment is not adequate in the deep lattice limit, where we use the GDTWA formalism, including the doublons and losses, to account for the observed dynamics.

The second fitting parameter of the theory is the quadratic Zeeman shift $q$. Since it arises  from tensorial light  shifts, it is expected to vary with lattice depth $V_0$, and while its precise value is not known \textit{ab initio}, experimental evidence \cite{Lepoutre2018b} suggests that its magnitude  is bounded by $\vert q/h \vert \leq 6$ Hz in the range of lattice depths considered.
The observed dynamics, most notably at short times, depends strongly on the sign and magnitude of $q$ (e.g.~see Eq.~(\ref{N0pert})).  
We have determined the values of $q$ that give the best fit to the initial dynamics. 
 The optimized values of $q$ are  shown in Fig.~\ref{fig:experiment_comp}(o) for the Gutzwiller and GDTWA methods. 

For $V_{0}<10E_r$ the optimal value of $q$ determined from the Gutzwiller method is within the expected range of values. However for deeper lattices a much larger $q$ value is required to achieve dynamics of a similar magnitude of that observed experimentally. This suggests that the Gutzwiller dynamics is failing in the deep lattice regime, consistent with our observations of the double-well dynamics. In fact, we know that quantum fluctuations, which play a key role  to drive the dynamics in  the Mott-insulator regime,  are not accounted for in the Gutzwiller model. Hence, the artificially large  $q$ value obtained when fitting Gutzwiller results to the experimental data is compensating  for the absence of these fluctuations in the simulation. 
For the GDTWA calculations a value of $q/h=2$ Hz  is found to provide good agreement for all lattice depths $>8E_r$.

In Fig.~\ref{fig:experiment_comp}(p) we show  the spatial distributions of the different levels  at $t=5$~ms  and  various lattice depths in the Gutzwiller model, which is a feature that has not been resolved in our experiment.
At low lattice depths  (see $V_0=2\,E_r$ result) the Gutzwiller method shows that the spin states  spatially separate in wavepackets that are appreciably narrower than the initial density distribution (black).
The spin-dependent transport is driven by the magnetic field gradient and causes the center of mass of the various wavepackets to undergo spatial oscillations. The packets tend to be well separated at the time when there is a large dip in the $N_0$ population in Fig.~\ref{fig:experiment_comp}(a) (and corresponding peak in the higher $|m|$-populations). This observation emphasizes the relevant role played by the interplay between the magnetic field gradient, which drives the spin transport, and the spin dependent contact interactions  which lead to  a  redistribution of the spin populations
as observed in the Gutzwiller predictions (see   Fig.~\ref{fig:experiment_comp}(a)-(d)). As the lattice depth increases spin transport is inhibited, and the wavepacket oscillations gradually become less visible. The Gutzwiller results for the density distributions at $V_0=8\,E_r$ reveals a small center of mass separation of the spin components  (here the wavepackets are broad and mostly overlapping).
The Gutzwiller solution predicts the system to become fully insulating, with one atom per site and all transport dynamics freezing out, as the lattice depth exceeds a critical value of $V_c\approx8E_r$. For $V_0>V_c$ the Gutzwiller model fails to capture the experimentally observed spin population dynamics.
On the contrary in this regime the GDTWA is able to reproduce very well the observed dynamics, even for $V_0$ as shallow as $V_c$, as shown in panels Fig. 3(h)-(n).

The sharp  change of behavior of the Gutzwiller dynamics as the system crosses the Mott insulator transition  is further illustrated in Fig~\ref{fig:withV0}(a)-(c), which shows the $m=0$ populations with lattice depth at three times. There we can observe that the Gutzwiller calculations show almost no dependence as a  function of lattice depth as soon as the Mott transition is reached. This can also be seen in Fig~\ref{fig:experiment_comp}, which shows that dynamics almost disappears in the Gutwiller approach above $V_c$.  Such a change in behavior around $V_c=8E_r$ is also observable in the experimental result but in a much less pronounced way. This is because, above the Mott transition, spin dynamics mainly occurs due to inter-site dipole-dipole interactions, and involves the growth of quantum correlations, a feature which can be reproduced by the GDTWA approach \cite{Lepoutre2018b} but not by the Gutzwiller simulations. As a consequence, there still exists significant dynamics at large lattice depths, which the Gutwiller ansatz fails to capture.

Although we only observe a gradual and smooth evolution of spin dynamics as the lattice spans the Mott to superfluid transition, it is worth pointing out that the observations are qualitatively different above and below the transition. Below the transition an oscillation is observed, and above the transition the population follows a monotonous evolution, with weak lattice depth dependence. To better quantify this claim, we plot in Fig. \ref{fig:withV0} (d) the time at which the $m=0$ fractional populations decay from their initial value to 0.25.
A clear difference is observed below and above the Mott transition. While the Gutzwiller model overestimates the amplitude of the oscillation for $V_0<V_c$ it correctly captures its shape. In particular, it is able to capture the slowing down of the dynamics which is experimentally observed as the lattice depth is raised, as highlighted in Fig. \ref{fig:withV0} (d). To physically understand this slowing down, it is worth referring to our recent results \cite{Lepoutre2018a}. Indeed, when the dipoles are tilted compared to the magnetic field axis by $\pi/2$, no spin dynamics can arise (either due to contact interactions or to dipole-dipole interactions) in absence of magnetic field gradients. To leave the initially polarized stationary state, a coupling between spin and motional degrees of freedom  is needed which is here provided by the gradient of the magnetic field. As shown in Fig.~\ref{fig:experiment_comp}(p), an important  effect of the lattice potential is to reduce the relative motion of the different spin states, therefore reducing the effect of the magnetic field gradient, and damping spin dynamics. 

On the other hand, only the GDTWA successfully captures the magnitude of the population decay for $V_0>V_c$. In fact, an unexpected outcome of the comparison between the experimental data and the GDTWA is that GDTWA reproduces well the dynamics even for lattice depths barely above the Mott transition. At large lattice depths, the disagreement of the experimental data with Gutzwiller results together with the good agreement with the GDTWA approach confirm that a new regime is reached, where spin dynamics is governed by different physical processes. As mentioned above, the large lattice depth regime is impacted by strong quantum correlations. We also observe in the simulations that, while spin dynamics is mostly driven by contact interactions at shallow lattice depths, it is almost entirely driven by dipole-dipole interactions at large lattice depths ($V_0>V_c$) .     

\begin{figure*}
	\includegraphics[width=0.9\textwidth]{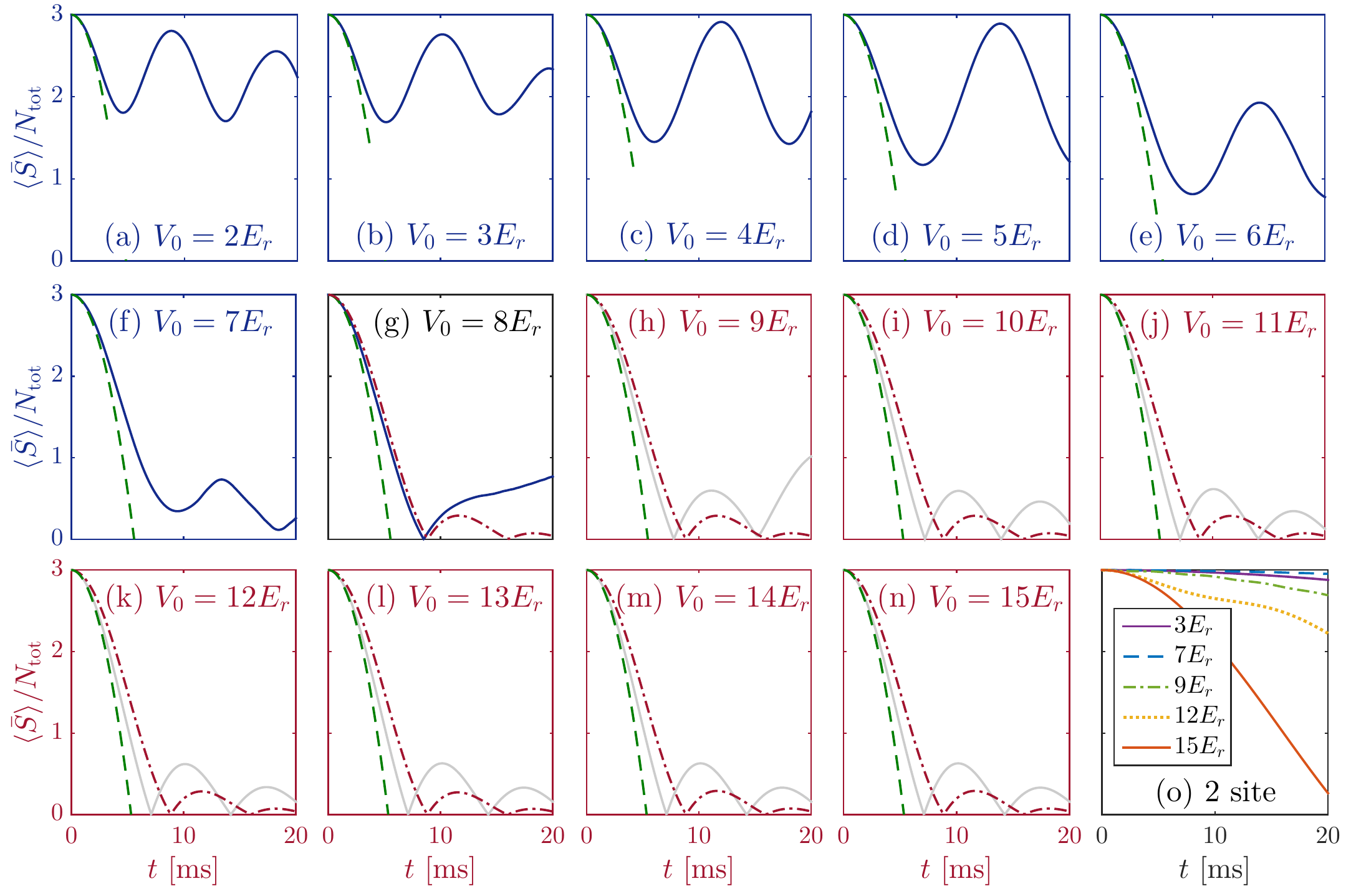}
	\caption{(a-n) Total spin as defined by Eq.~(\ref{eq:Stot_def}) for a range of lattice depths. The blue solid line indicates the total spin using the Gutzwiller model, with the results beyond the lattice depth $V_c$ that the Gutzwiller model is considered reliable shown in gray. The red dot-dashed line indicates the GDTWA result and green dashed line gives the short time second order expansion, found by summing Eq.~(\ref{eq:Stot}) over all pairs of atoms. (o) Total spin in a double-well system (see Fig.~\ref{fig:shorttime}) using the exact method with different lattice depths} 
	\label{fig:tot_spin}
\end{figure*}

\section{Total Magnetization and Gap Protection}

The dynamical evolution of the populations of the Zeeman levels, as presented in Fig.~\ref{fig:experiment_comp}(a)-(n), varies smoothly when crossing the underlying superfluid to Mott transition. 
In order to better understand 
the spin  dynamics both  in the Superfluid and Mott regimes, we now study  the dynamics of the collective spin length, corresponding to the total magnetization, which we find better reveals the underlying abrupt transition. It is given by:

\begin{equation}
\label{eq:Stot_def}
\langle \bar S \rangle =\sqrt{\langle { \hat S^x} \rangle^2+\langle { \hat S^y} \rangle^2+\langle { \hat S^z} \rangle^2 },
\end{equation} Here $\hat S^{x,y,z}=\sum_i\sum_{m,n} S_{mn}^\alpha \hat{a}^\dagger _{m,i} \hat{a}_{n,i}$ are collective spin observables. For the case of an initial state following a $\pi/2$ rotation, the quantity  $\langle { \hat S^z} \rangle$ which is conserved during the dynamics is equal to zero, and therefore $\langle \bar S \rangle =\sqrt{\langle { \hat S^x} \rangle^2+\langle { \hat S^y} \rangle^2}$.

In Fig.~\ref{fig:tot_spin} we plot the dynamics of the  collective spin length obtained  by the Gutzwiller predictions ($V_0<8 E_r$) and GDTWA ($V_0\geq 8E_r$). The dashed green line also shows the short time dynamics obtained from perturbation  theory of the double-well system  up to terms  quadratic in time, $O(t^2)$, summed over pairs of atoms at neighbouring lattice sites: 

\begin{align}
\label{eq:Stot}
\langle \bar S \rangle \sim \sum _i N^i \bigg(3-\frac{3t^{2}}{2}\bigg[&5q^{2}+(\gamma b(\mathbf{r}_{i}))^{2}\\
\nonumber &+\frac{27}{2}\sum_{j^{(i)}}\left(U_{dd}(\mathbf{r}_i-\mathbf{r}_j)\right)^2\bigg]\bigg),
\end{align}
where $N^i$ is the population at site $i$ and the sum over $ j^{(i)}$ includes all of site $i$'s nearest neighbours. This is dominated by the large magnetic gradient term, and is therefore largely independent of lattice depth. 

The collective spin length provides direct information of the spin coherence and it is experimentally accessible in a Ramsey sequence \cite{Yan2013,Hazzard2014b,Lepoutre2018a} performed by applying a $\pi/2$ pulse after the free evolution before measuring the population. 
Two leading processes are expected to generate magnetization decay: the field gradient and the interactions. While the magnetic gradient generates single particle dephasing since it causes different  lattice sites to precess at slightly different rates, interactions entangle the spins leading to a loss of information when one traces out over a part of the total system as one does when computing  local observables such $\langle \hat s_i^{x,y}\rangle$. The quadratic Zeeman term can also lead to magnetization decay in this case due to the development of intra-spin correlations, i.e. correlations between the individual electrons inside each atom \cite{Polzik2008}.

While generically, interactions and inhomogeneities both can lead to magnetization decay, the spin dependent interactions proportional to $U_1$ in Eq. (\ref{Ham})  counter-intuitively  can favor spin alignment  for weakly interacting atoms.  This is because these interactions open a gap in the energy spectra that can suppress dephasing processes  as experimentally demonstrated in recent work \cite{Lepoutre2018a, Norcia2018,Smale2019}. Signatures of the gap protection can be observed in Gutzwiller predictions for  shallow lattices  Fig.~\ref{fig:tot_spin} (a-f). The protection is present when there are more than one atom per lattice site and when the $U_1$ term can lock the spins favoring alignment. This manifests as an  oscillatory behavior in the collective spin length instead of rapid decay.  The protection however enters as a higher order process and it is not observable in the perturbative analysis that neglects third and higher  order terms in time. The perturbative analysis however does provide a relatively good description of the short time dynamics. 

In the Mott regime on the other hand interactions and magnetic field gradient cooperate and both lead to a fast decay of the contrast as can be observed in the GDTWA simulations  Fig.~\ref{fig:tot_spin} (h-n). Similar behavior is observed in the exact solution of a double well system Fig.~\ref{fig:tot_spin} (o), validating the behavior observed in the many-body system.

The relatively sharp transition between oscillatory and overdamped  behavior around the critical point  ($V_{0}=7E_{r}$ to $V_{0}=8E_{r}$) might be overestimated in the Gutzwiller approximation but might survive in the full quantum system. Experimental measurements of the contrast will be needed to test if this is the case.





\section{Conclusions}

 In this work we presented extensive theoretical and experimental comparisons of the dynamics of itinerant spin-3 Cr atoms in a 3D optical lattice and subject to harmonic trapping along all three directions.
The microscopic Hamiltonian governing the dynamics of the system is complex and features single particle motion, as well as contact, and long-range dipolar interactions. Exact modeling of the experimental dynamics exceeds the capabilities of classical computation, so
to compare to our experimental results, we have developed a variety of approximate models. First, we studied  the exact population dynamics of the Zeeman levels of a two-site system and compared our results to a Gutzwiller treatment, as well as a short-time perturbative treatment. We thus demonstrated that for shallow lattice depths below the superfluid-insulator phase transition at $V_c$, where a significant portion of the sites are doubly occupied, the Gutzwiller description provides a good qualitative description of the exact dynamics. However, we found that the Gutzwiller description fails reproducing the exact dynamics for large lattice depths, above the Mott transition. For large lattice depths, on the other hand, we have used an effective spin model description (GDTWA) whose short-time dynamics was shown in~\cite{Lepoutre2018b}  to match the exact solution for large enough lattice depths, where tunneling and double occupancy are suppressed. 

Armed with the above intuition, we applied the Gutzwiller approximation to lattice depths $V_0<V_c$ and the GDTWA method, which incorporates the effect of quantum fluctuations, to $V_0>V_c$ to benchmark the experimental observations. We observed qualitative agreement between the experimental results and the theoretical studies, which confirms that these approximate methods can be trusted in their respective domain of validity. In turn, the comparison with these two models provides unique physical insights regarding the physics at play for different lattice depths. While that for $V_0>V_c$ dipolar interactions and quantum effects play an essential role in the observed dynamics, we found that for $V_0<V_c$ transport and contact interactions both play an essential role (while intersite quantum correlations can then be neglected). Our analysis thus shows that the two different regimes of low and high lattice depths are qualitatively different. This can indeed be seen by contrasting the behavior of spin dynamics in these two regimes: while spin dynamics is oscillatory and lattice depth dependent for $V_0<V_c$, the behavior is mostly monotonous and lattice depth independent for $V_0 > V_c$. The cross-over between these two behaviors is however smooth, and does not reveal the sharp underlying Mott to superfluid transition. Therefore, we propose an experimental measurement of the spin-length which can be readily implemented experimentally, as it should display a more pronounced change in behaviour as the system crosses from the superfluid to Mott insulator regime. 
\begin{acknowledgments}
 We  thank  Johannes Schachenmayer for useful discussions and Asier Pineiro and Andrew Wilson for reviewing the manuscript. {\bf Funding:}  A.M.R is supported by the AFOSR grant FA9550-18-1-0319 and its MURI Initiative, by the DARPA and ARO grant W911NF-16-1-0576, the DARPA DRINQs grant, the ARO single investigator award W911NF-19-1-0210,  the NSF PHY1820885, NSF JILA-PFC PHY-1734006 grants, and by NIST. The Villetaneuse group acknowledges financial support from Conseil R\'egional d'Ile-de-France under DIM Nano-K / IFRAF, CNRS, Minist\`ere de l'Enseignement Sup\'erieur et de la Recherche within CPER Contract, Universit\'e Sorbonne Paris Cit\'e (USPC), and the Indo-French Centre for the Promotion of Advanced Research - CEFIPRA under the LORIC5404-1 contract. A.S.N was supported by the Australian Research Council Centre of Excellence for Engineered Quantum Systems (project number CE170100009), and funded by the Australian Government.
 B.Z. is supported by the NSF through a grant to ITAMP.
\end{acknowledgments}




\appendix
\section{Numerical Methods} \label{app:nummethods}

The Gutzwiller dynamical equations can be obtained by variationally
minimizing $\langle\psi_{G}|i\hbar\frac{\partial}{\partial t}-\hat{H}|\psi_{G}\rangle$.
This yields a set of nonlinear differential equations for the evolution
of the Gutzwiller coefficients $\{f_{\mathbf{N}}^{i}\}$, i.e.
\begin{equation}
i\hbar\frac{\partial f_{\mathbf{N}}^{j}}{\partial t}=F_{\mathbf{N}}^{j}(\{f_{\mathbf{N}}^{i}\}),\label{eq:Gutz}
\end{equation}
where $F_{\mathbf{N}}^{j}(\{f_{\mathbf{N}}^{i}\})= _{j}\langle\mathbf{N}|\hat{H}_{G}^{j}|\psi_{G}(t)\rangle.$
The form of the $F_{\mathbf{N}}^{j}(\{f_{\mathbf{N}}^{i}\})$ is analytically
cumbersome (e.g. see \cite{Asaoka2016a} for the spin-1 case), but is easily evaluated numerically
by taking expectations of the various operators terms appearing in
$\hat{H}_{G}^{j}$ in terms of the Gutzwiller coefficients. From the
initial condition the system of equations (\ref{eq:Gutz}) is evolved
using an adaptive step Runge-Kutta method, with the tolerance set
sufficiently low that the solution converges.

\section{Parameters }\label{app:parameters}
The coefficients for the contact terms are given by Ref.~\cite{Kawaguchi2012a} as
\begin{equation}
\tilde{c}_{0}=c_{0}-\frac{c_{3}}{7},\quad\tilde{c}_{1}=c_{1}-\frac{5c_{3}}{84}\quad\tilde{c}_{2}=c_{2}-\frac{5c_{3}}{3},\quad\tilde{c}_{3}=\frac{c_{3}}{126}
\end{equation}
where  $c_{0}=71g_{c}a_{B}$, $c_{1}=3g_{c}a_{B}$, $c_{2}=-15g_{c}a_{B}$, and $c_{3}=-46g_{c}a_{B}$ with $g_{c}=4\pi\hbar^{2}/M$ and $a_{B}$ is the Bohr radius. This gives $U_{n}=\tilde{c}_{n}\int d\mathbf{r}|\text{w}(\mathbf{r})|^{4}$ . The contact interaction also contains a $U_2$ and $U_3$ term
\begin{align}
	C^{mm'nn'}=&  U_{0}\delta_{m,n}\delta_{m',n'}+U_{1}\sum_{\alpha}S_{mn}^{\alpha}S_{m'n'}^{\alpha}\\
	 +&\frac{U_{2}}{7}(-1)^{m+n}\delta_{m,-m'}\delta_{n',-n}\nonumber\\
	 +&\frac{U_{3}}{2}\left[\sum_{\alpha\beta}\left(S^{\alpha}S^{\beta}\right)_{mn}\left(S^{\alpha}S^{\beta}+S^{\beta}S^{\alpha}\right)_{m'n'}\right]\nonumber
\end{align}
which are included in our model, but have negligible effect for the states accessible during the dynamics in consideration compared to the larger $U_0$ and $U_1$ contributions. 

\begin{figure}
	\includegraphics[width=0.4\textwidth]{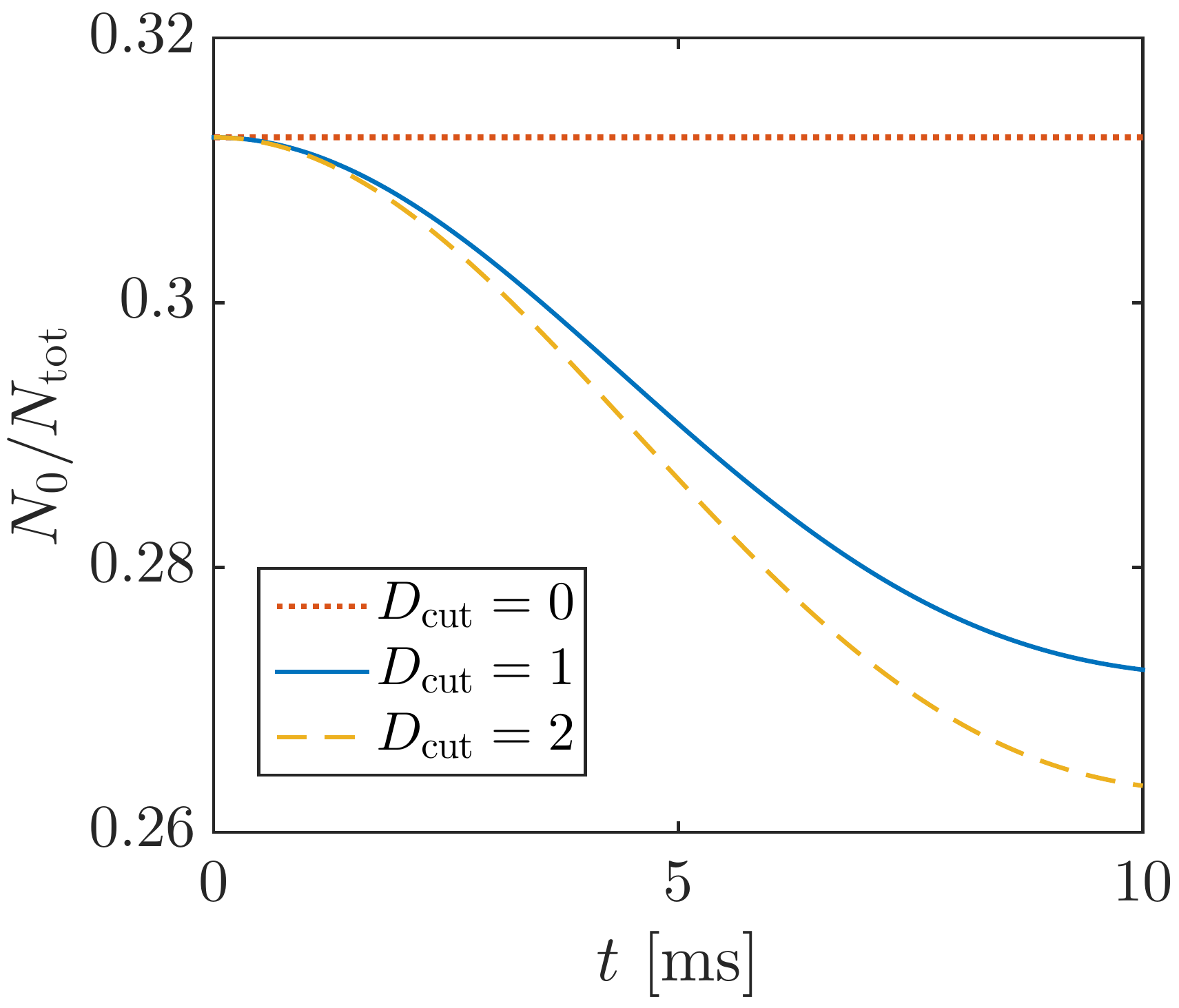}
	
	\caption{Comparison of DDI cutoff approximations in the Gutzwiller model for $V_0=15E_r$ and $q/h=15$ Hz. When $D_{\rm{cut}}=0$ only onsite DDIs are included, when  $D_{\rm{cut}}=1$ nearest neighbours (including diagonal nearest neighbours) and included and when  $D_{\rm{cut}}=2$ second nearest neighbours are also included. We see that the contribution from second nearest neighbours is small compared to the nearest neighbours. As intersite DDIs are most important in the deep lattice limit, this conclusion will hold for all lattice depths.}
    \label{fig:DDI_cutoff}
\end{figure}

In this paper, with the exception of the GDTWA treatment, we use the nearest-neighbour as the cut-off for the dipolar interactions. This is justified as in shallower lattices where doubly occupied sites exist, the on-site dipolar interactions are dominant. Generally, we find that for the range of lattice depths considered in this paper, the effect of second nearest neighbour interactions is small compared to that of nearest neighbour interactions. We provide a comparison between these terms in Fig.~\ref{fig:DDI_cutoff}. 

\begin{figure*}
	\includegraphics[width=0.9\textwidth]{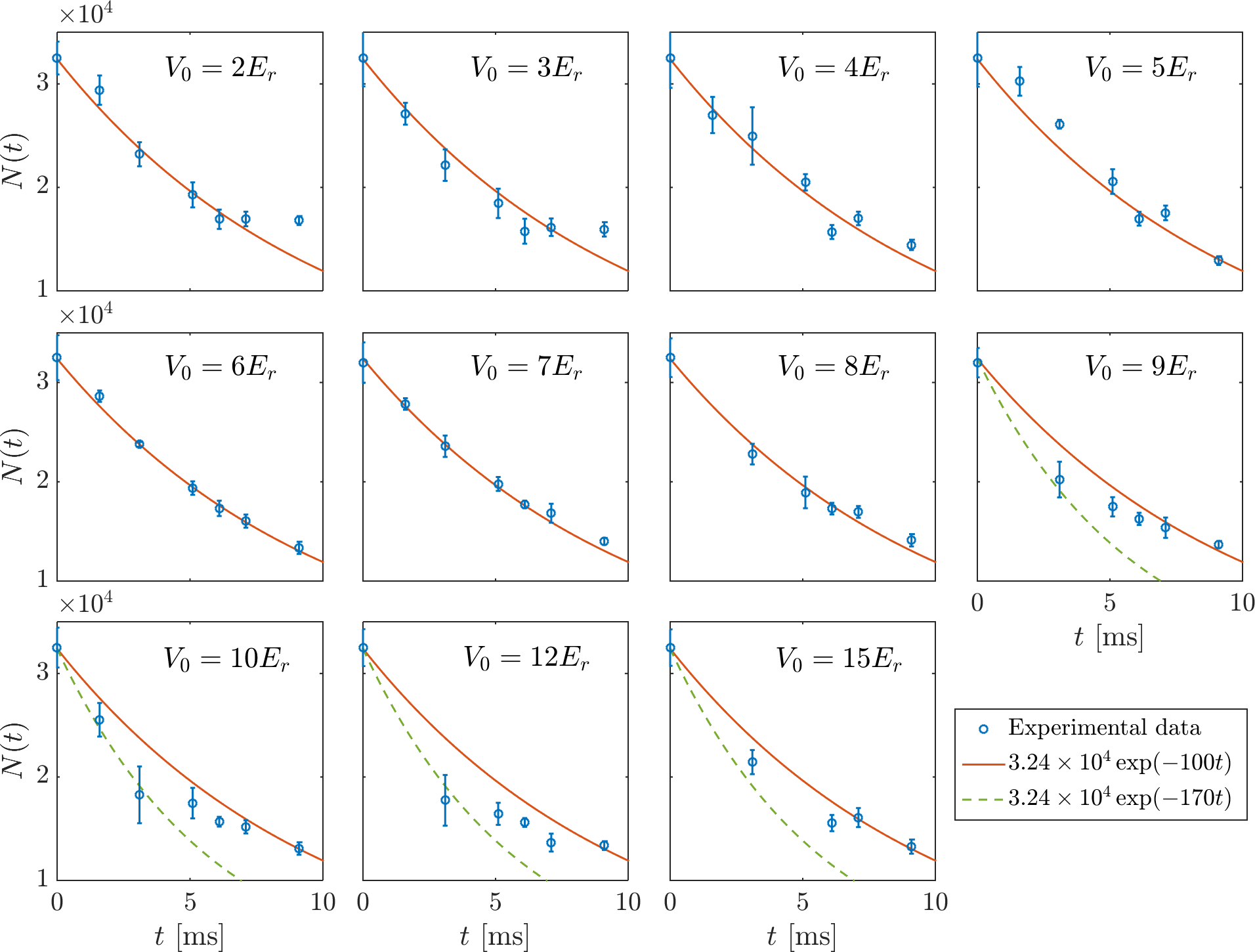}
	
	\caption{Experimental values for the total number of atoms in the lattice, along side approximate exponential decay with $\gamma=100\,s^{-1}$ and $170\,s^{-1}$.}
    \label{fig:Atomloss}
\end{figure*}

Atom loss is significant in the experiment, and in Fig. \ref{fig:Atomloss} we show the experimental result for the population over time, along with the $N(t)= N(0)\exp(-\gamma t)$ approximation for two values of $\gamma$. At low lattice depths $\gamma=100\,s^{-1}$ fits the experimental data well, while at larger lattice depths there is an initial fast decay of $\gamma=170\,s^{-1}$, which then slows and returns to the $\gamma=100\,s^{-1}$ value by $10$ ms.

\section{Perturbative Expressions for Substate Population Dynamics }\label{app:pertexp}

\subsection{Short-time dynamics at $O(t^2)$}
In the main text, Eq.~(\ref{N0pert}) provides a representative expression for the perturbative (short-time) dynamics of the substate populations for an initial state of singlons. Here we provide the rest of the expressions which take a similar form to those presented in the main text.  More generally we can write
\begin{align}
	\vert\psi(0)\rangle=&\sum_{m,n}f_{m}f_{n}\biggl[p_{1}\hat a_{m, 1}^\dagger \hat a_{n, 2}^\dagger \\
	 +&\frac{p_{2}}{2}\sqrt{1+\delta_{m,n}}\left(\hat a_{m,1}^\dagger \hat  a_{n,1}^\dagger + \hat a_{m,2}^\dagger \hat a_{n,2}^\dagger \right)\biggr]\nonumber
	\times \vert 0 \rangle_1 \vert 0 \rangle_2
\end{align}
where $p_{1}$ is the amplitude for the state with one atom per
site and $p_{2}=\sqrt{1-p_{1}^{2}}$ is the amplitude of the state
with two atoms on one site and zero on the other, both of which are
chosen to be real. To second order in $t$ we find the substate population dynamics is given by,
\begin{align} 
N_{m}^{(2)}(t)&=p_{1}^{2}N^{(2)}_{m, \lbrace 1,1 \rbrace} (t)+p_{2}^{2}N^{(2)}_{m,\lbrace 2,0 \rbrace}(t)
\end{align}
where we have used $N^{(2)}_{m, \lbrace 1,1\rbrace}$ and $N^{(2)}_{m, \lbrace 2,0\rbrace}$ to indicate the expectation value of $\hat N_m$ for pairs of sites with singlons ($n_s$) and the number of atoms in a doublon ($n_d$) configuration, respectively. These are given by 

\begin{align} 
N_{0, \lbrace 1,1 \rbrace}^{(2)} (t)&=\frac{ 5}{8}-\frac{n_s\left(n_s-1\right)}{2}\frac{45}{16}t^{2}U_{dd}^{1}\left(q+\frac{3}{4}U_{dd}^{1}\right),
\end{align}
\begin{align}
N_{0,\lbrace 2,0 \rbrace}^{(2)}(t)=\frac{5}{8}-&\frac{n_d\left(n_d-1\right)}{2}\frac{45}{8}t^{2}\nonumber\\
&\quad \times\left(U_{1}+\frac{1}{2}U_{dd}^{0}\right)\left(q+\frac{3}{4}U_{dd}^{0}\right),
\end{align}
\begin{align}
N_{\pm1, \lbrace 1,1 \rbrace}^{(2)} (t)&=\frac{15}{32}-\frac{ n_s(n_s-1)}{2}\frac{45}{64}t^{2}U_{dd}^{1}\left(q+\frac{3}{4}U_{dd}^{1}\right),
\end{align}
\begin{align}
N_{\pm1,\lbrace 2,0 \rbrace}^{(2)}(t)=\frac{15}{32}-&\frac{n_d(n_d-1)}{2}\frac{45}{32}t^{2}\nonumber\\
&\quad\times\left(U_{1}+\frac{1}{2}U_{dd}^{0}\right)\left(q+\frac{3}{4}U_{dd}^{0}\right),
\end{align}
\begin{align}
N_{\pm2, \lbrace 1,1 \rbrace}^{(2)} (t)&=\frac{3}{16}+\frac{ n_s(n_s-1)}{2}\frac{45}{32}t^{2}U_{dd}^{1}\left(q+\frac{3}{4}U_{dd}^{1}\right),
\end{align}
\begin{align}
N_{\pm2,\lbrace 2,0 \rbrace}^{(2)}(t)=\frac{3}{16}+&\frac{n_d(n_d-1)}{2}\frac{45}{16}t^{2}\nonumber\\
&\quad\times\left(U_{1}+\frac{1}{2}U_{dd}^{0}\right)\left(q+\frac{3}{4}U_{dd}^{0}\right),
\end{align}
\begin{align}
N_{\pm3, \lbrace 1,1 \rbrace}^{(2)} (t)&=\frac{1}{32}+\frac{ n_s(n_s-1)}{2}\frac{45}{64}t^{2}U_{dd}^{1}\left(q+\frac{3}{4}U_{dd}^{1}\right),
\end{align}
\begin{align}
 N_{\pm3,\lbrace 2,0\rbrace}^{(2)}(t)=\frac{1}{32}+&\frac{n_d(n_d-1)}{2}\frac{45}{32}t^{2}\nonumber\\
 &\quad\times\left(U_{1}+\frac{1}{2}U_{dd}^{0}\right)\left(q+\frac{3}{4}U_{dd}^{0}\right),
\end{align}
 where $U_{dd}^0=U_{dd}(\mathbf{0})$ is the onsite DDI and $U_{dd}^1$ is the DDI to the neighbouring site. In this derivation we  neglected the $U_2$ and $U_3$ terms as their contribution is negligible. 
\subsection{Short-time dynamics at $O(t^4)$}\label{app:4thorder}

In a double-well system, the $t^4$ contribution to the short-time dynamics is given by
\begin{equation}
N_{m}^{(4)}(t)=N_{m}^{(2)}(t)+\frac{t^{4}}{24}\left(p_{1}^{2}A^m_{1}+p_{2}^{2}A^m_{2}+p_{1}p_{2}A^m_{12}\right),
\end{equation}
where $A^m_1,A^m_{2}$, and $A^m_{12}$ are the terms due to state with only singlons, only doublons, and the superposition of the two, respectively. Setting $m=0$, we find
\begin{widetext}
\begin{align}
A^0_{1}= & \frac{45}{8}U_{dd}^{1}\left[B^{2}(3q-2U_{dd}^{1})+4(q+\frac{3}{4}U_{dd}^{1})\left(4q^{2}+\frac{21}{4}qU_{dd}^{1}+\frac{37}{4}\left(U_{dd}^{1}\right)^{2}\right)\right]\nonumber \\
 & -135J^{2}\left[(U_{1}+\frac{1}{2}U_{dd}^{0})(2q+\frac{3}{4}U_{dd}^{0})+\frac{1}{4}(-4q+3U_{1}+U_{dd}^{0}-\frac{5}{2}U_{dd}^{1})U_{dd}^{1}\right],\\
A^0_{2}= & \frac{45}{4}(U_{1}+\frac{1}{2}U_{dd}^{0})(q+\frac{3}{4}U_{dd}^{0})\left[2q\left(8q-3U_{1}+\frac{21}{2}U_{dd}^{0}\right)+121U_{1}^{2}+\frac{1}{2}U_{dd}^{0}\left(233U_{1}+74U_{dd}^{0}\right)\right]\nonumber \\
 & +135J^{2}\left[(U_{1}+\frac{1}{2}U_{dd}^{0})(2q+\frac{5}{4}U_{dd}^{0})-\frac{1}{4}(4q-U_{1}+U_{dd}^{0}+\frac{3}{2}U_{dd}^{1})U_{dd}^{1}\right],
\end{align}
\end{widetext}
where we have used $B\equiv\gamma\left[ b(\mathbf{r}_1)- b(\mathbf{r}_2)\right]$ for two nearest neighbouring sites at  $\mathbf r_1$ and $\mathbf r_2$. If the double well starts in a superposition of $\lbrace 1,1 \rbrace$ (singlons) and $\lbrace 2, 0 \rbrace$ (doublon-hole) states then we get an additional contribution to the fourth order dynamics given by, 
\begin{widetext}
\begin{align}
A^0_{12}= & \frac{45}{4}J\bigg[-B^{2}\left(U_{1}+\frac{1}{2}U_{dd}^{0}-U_{dd}^{1}\right)+3U_{dd}^{0}\left(U_{1}+\frac{1}{2}U_{dd}^{0}\right)\left(2U_{0}+7U_{1}+8U_{dd}^{0}\right)\nonumber\\
 & +q\left(6U_{0}+43U_{1}+35\left(U_{dd}^{0}-U_{dd}^{1}\right)\right)\left(2U_{1}+U_{dd}^{0}-U_{dd}^{1}\right)\nonumber\\
 &+\frac{3}{2}U_{dd}^{1}\left(2U_{0}U_{1}+29U_{1}^{2}+3U_{1}U_{dd}^{0}-8\left(U_{dd}^{0}\right)^{2}\right) -3\left(U_{0}+13U_{1}+4\left(U_{dd}^{0}-U_{dd}^{1}\right)\right)\left(U_{dd}^{1}\right)^{2}\bigg].
\end{align}
\end{widetext}
Similarly for the $m\pm1$ we find
\begin{align}
A^{\pm1}_{1} = \frac{1}{4}A^{0}_{1}+\frac{135}{4}U_{dd}^1\left(q-\frac{7}{2}U_{dd}^1\right)\left(q+\frac{3}{4}U_{dd}^1\right)^2,
\end{align}\begin{align}
A^{\pm1}_{2} = \frac{1}{4}A^{0}_{2}+\frac{135}{4}&\left(2q-17U_1-7U_{dd}^0\right)\nonumber\\
&\left(U_1+\frac{1}{2}U_{dd}^0\right)\left(q+\frac{3}{4}U_{dd}^0\right)^2,
\end{align}
and \begin{align}
A^{\pm1}_{12} = \frac{1}{4}A^{0}_{12}.
\end{align}
The remaining terms are given by $A_n^{\pm2}=-\frac{1}{2}A_n^0$ and $A_n^{\pm3}=-A_n^{\pm1}$.
\subsection{Divergence Time Scales}\label{app:timescale}
In the inset of Fig. \ref{fig:shorttime}(b) we see that the second order results including doublons rapidly diverge from the exact simulation. To predict the timescales over which this divergence occurs, we consider the largest term in the doublon expansion
\begin{align}
   \Delta N_{0}^{(2)}=-\frac{45(2\pi)^{2}t^{2}}{8}U_{1}(q+0.75U_{dd}^{0}).
\end{align}
The $y$ axis in the inset covers $\Delta N_{0}=6\times10^{-5}$, and we find  $\Delta N_{0}^{(2)}$ changes by this amount in $t=0.03$ ms, which agrees well with the figure. Divergence on a similar order of magnitude to the main figure (i.e. $\Delta N_{0}=0.15$), occurs when the higher order terms become important. Considering the largest $4^{th}$ order term, which depends on tunnelling, we get
\begin{align}
\Delta N_{0}^{(4)}=\frac{135(2\pi)^{4}t^{4}}{24}J^{2}U_{1}(2q+\frac{5}{4}U_{dd}^{0}+\frac{1}{4}U_{dd}^{1}).
\end{align}
Note that this predicts $\Delta N_{0}^{(4)}=0.15$ at $t=0.29$ ms,  which indicates that the $2^{nd}$ order results will breakdown for time scales on the order of $t=0.29$ ms.

%

\end{document}